\pdfoutput=1

\documentclass[11pt]{article}
\usepackage{graphicx}
\usepackage{amsmath}
\usepackage{amsfonts}
\usepackage{amssymb}
\usepackage{color}

\newcommand{\be}{\begin{equation}}
\newcommand{\ee}{\end{equation}}
\newcommand{\ba}{\begin{eqnarray}}
\newcommand{\ea}{\end{eqnarray}}
\newcommand{\nn}{\nonumber}
\newcommand{\barr}{\begin{array}}
\newcommand{\earr}{\end{array}}
\newcommand{\eqdef}{\stackrel{\rm def}{=}}

\newcommand\lsim{\mathrel{\rlap{\lower4pt\hbox{\hskip1pt$\sim$}}
        \raise1pt\hbox{$<$}}}
\newcommand\gsim{\mathrel{\rlap{\lower4pt\hbox{\hskip1pt$\sim$}}
        \raise1pt\hbox{$>$}}}

\def\n{{\bf \widehat n}}
\def\threej#1#2#3#4#5#6{\left( \begin{array}{ccc} #1 & #2 & #3 \\ #4 & #5 & #6 \end{array} \right) }
\def\fsky{f_{\rm sky}}
\def\ellmin{\ell_{\rm min}}
\def\ellmax{\ell_{\rm max}}
\def\zmax{z_{\rm max}}
\def\bigoh{{\mathcal O}}

\begin{document}

\begin{titlepage}
\setcounter{page}{1} \baselineskip=15.5pt \thispagestyle{empty}

\bigskip\
\begin{center}
{\Large \bf Delensing CMB Polarization with External Datasets}
\end{center}
\vspace{0.5cm}
\begin{center}
{\fontsize{14}{30}\selectfont Kendrick M.~Smith$^1$, Duncan Hanson$^{2}$, Marilena LoVerde$^3$,} \\
\vskip 4pt
{\fontsize{14}{30}\selectfont Christopher M.~Hirata$^4$ and Oliver Zahn$^{5,6}$}
\end{center}

\begin{center}
\vskip 8pt
\textsl{${}^1$ Princeton University Observatory, Peyton Hall, Ivy Lane, Princeton, NJ 08544 USA}
\vskip 4pt
\textsl{${}^2$ Institute of Astronomy and Kavli Institute for Cosmology} \\
\textsl{University of Cambridge, Madingley Road, Cambridge CB3 0HA, UK}
\vskip 4pt
\textsl{${}^3$ Institute for Advanced Study, Einstein Drive, Princeton, NJ 08540, USA}
\vskip 4pt
\textsl{${}^4$ California Institute of Technology, Mail Code 350-17, Pasadena, CA 91125}
\vskip 4pt
\textsl{${}^5$ Berkeley Center for Cosmological Physics, Department of Physics} \\
\textsl{University of California, Berkeley, CA 94720, USA}
\vskip 4pt
\textsl{${}^6$ Lawrence Berkeley National Labs, University of California, Berkeley, CA 94720, USA}
\end{center}

\vspace{1cm}

\hrule \vspace{0.3cm}
{ \noindent \textbf{Abstract} \\[0.2cm]
\noindent
One of the primary scientific targets of current and future CMB polarization experiments is the search
for a stochastic background of gravity waves in the early universe.  As instrumental sensitivity improves,
the limiting factor will eventually be B-mode power generated by gravitational lensing, which can be
removed through use of so-called ``delensing'' algorithms.  We forecast prospects for delensing
using lensing maps which are obtained externally to CMB polarization: either from large-scale structure
observations, or from high-resolution maps of CMB temperature.  We conclude that the forecasts in either
case are not encouraging, and that significantly delensing large-scale CMB polarization
requires high-resolution polarization maps with sufficient sensitivity to measure the lensing B-mode.
We also present a simple formalism for including delensing in CMB forecasts which is computationally
fast and agrees well with Monte Carlos.}
\vspace{0.3cm}
\hrule

\vspace{0.6cm}
\end{titlepage}

\section{Introduction}
\label{sec:intro}

The last two decades of cosmological observations have resulted in a model which is consistent
with a wide variety of datasets (e.g.~\cite{Komatsu:2010fb,Percival:BAO,Riess:H,Kessler:SN,Vikhlinin:Xray})
to an impressive level of precision.
The observations are consistent with a flat $\Lambda$CDM expansion history, and 
Gaussian adiabatic initial conditions with a power spectrum which is slightly redder
than scale-invariant ($n_s = 0.963 \pm 0.012$).

A major observational frontier in the near future will be the search for a stochastic background of gravity
waves on cosmological scales, parameterized by the tensor-to-scalar ratio $r$.
In single-field slow-roll inflation, $r$ is related to the total field variation of the inflaton in Planck units
($\Delta\phi = (M_{Pl}/8\sqrt{\pi}) \int r^{1/2} dN$); 
``large-field'' models with Planck-scale field excursions correspond to $r \gsim 10^{-2}$, whereas ``small-field''
models can have arbitrarily small values of $r$ \cite{Lyth:1996im}.
The simplest models of inflation, such as a power-law slow-roll potential, have $r\sim 10^{-2}$, and it has
been argued that obtaining $r$ smaller than this requires fine-tuning \cite{Boyle:r}.
However, there are also models with a microphysical reason for requiring small $r$, such as the string-inspired
KKLMMT model \cite{KKLMMT}, so this is not a firm prediction.
Increasingly precise constraints on $r$ in the next few years will sharpen this picture considerably.

% ekpyrosis here?
% DBI inflation here?

On the observational front, current upper limits are $r \le 0.36$ (95\% C.L.) from large-scale CMB
temperature \cite{Komatsu:2010fb}, or $r \le 0.72$ (95\% C.L.) from large-scale CMB B-mode polarization 
\cite{Chiang:2009xsa}.
Because the large-scale CMB temperature contains large primary contributions (i.e.~Sachs-Wolfe
and Integrated Sachs-Wolfe anisotropy), the temperature constraint on $r$ is limited by sample 
variance (and uncertainty in other cosmological parameters) and cannot be qualitatively improved.
In contrast, large-scale B-modes are only generated by gravity waves (or other non-scalar sources,
in models which contain them) or nonlinear effects \cite{Kamionkowski:1996ks,Zaldarriaga:1996xe},
and the polarization constraint on $r$ is currently noise-limited.
As instrumental sensitivity improves, the statistical error will vary roughly as $\sigma(r) \propto 1/T$,
where $T$ is the sensitivity of the experiment in detector-hours.
Since detector sensitivities and array sizes of CMB polarization experiments are currently in a state of rapid
development, polarization experiments will give the best constraints on $r$ in the near future, and should soon 
probe the $r \sim 10^{-3}$ to $10^{-2}$ range where qualitative classes of early universe models can be ruled out.

In the limit of small instrumental noise, the largest ``guaranteed'' source of B-mode power on large angular
scales is gravitational lensing by large-scale structure, which converts the E-modes generated during recombination
to a mixture of E and B \cite{Zaldarriaga:1998ar,Hu:2000ee}.
The lensed B-mode power spectrum $C_\ell^{B_{\rm len}}$ is roughly constant on large angular scales, and can
thus be interpreted as an extra source of white noise, whose amplitude is found to be 4.4 $\mu$K-arcmin for
the fiducial cosmology used in this paper.
When instrumental noise levels fall below this level, expriments will be {\em lensing-limited}: the limiting
factor in constraining $r$ will be the lensing B-mode, which acts as the dominant source of noise.

``Delensing'' algorithms have been proposed \cite{Knox:2002pe,Kesden:2002ku,Seljak:2003pn} which statistically separate the gravity wave and
lensing B-mode signals, offering the prospect for reducing the effective noise level below the 4.4 $\mu$K-arcmin
floor due to gravitational lensing.
These algorithms are based on the idea that, given a noisy observation of the CMB E-mode polarization and
an estimate for the CMB lensing deflection field, one can form an estimate for the large-scale lensing B-mode and 
subtract it from the observed B-mode to reduce the level of lensing contamination.
The required estimate for the deflection field is usually obtained ``internally'', using a lens reconstruction
estimator \cite{Benabed:2000jt,Guzik:2000ju,Hu:2001kj,Hirata:2003ka} which estimates the lensing deflection field 
directly from observations of the lensed CMB polarization.
High angular resolution is required for delensing (even if one is only interested in delensing B-modes on large
angular scales), since the large-scale lensing B-mode receives contributions from E-mode and deflection power on
small scales.
In this way, after B-mode experiments become lensing-limited, the small and large scales will be intimately linked:
the effective noise level for the large-scale B-mode will be determined by the fidelity with which the small-scale
polarization is measured, and used to estimate the deflection field.

The purpose of this paper is twofold: first, we present an approximate scheme for doing forecasts which include
delensing.  This scheme is easy to implement, computationally fast (a few CPU-seconds are needed per forecast)
and we find that it agrees well with Monte Carlo based results in the literature;
second, we investigate whether delensing can be made more effective by making use of lensing maps which
are obtained externally from CMB polarization.
Small-scale CMB {\em temperature} maps will soon be available (e.g.~from the Planck satellite \cite{BLUEBOOK}, or 
from ground-based experiments such as ACT \cite{Das:2010ga} or SPT \cite{Lueker:2009rx}) which can be used to reconstruct the CMB
lensing map with reasonable signal-to-noise.
Approximate CMB lensing maps can also be obtained using observations of large-scale structure (e.g.~cosmic shear), but
such maps will not have precisely the same redshift weighting as the CMB, where the contribution from large-scale structure
at high redshifts is non-negligible.
Do either of these sources of lensing information suffice (or help significantly) to delens large-scale CMB polarization?

Throughout this paper, we use the WMAP7+BAO+H$_0$ cosmology from \cite{Larson:2010gs}, with parameters $\Omega_bh^2 = 0.227$,
$\Omega_ch^2 = 0.111$, $h=0.714$, $n_s = 0.969$, $\Delta^2_{\mathcal R}(k_0) = 2.38 \times 10^{-9}$ at wavenumber
$k_0 = 0.002$ Mpc$^{-1}$, and $\tau = 0.086$.
All power spectra were calculated using CAMB \cite{Lewis:1999bs}, using HALOFIT \cite{Smith:2002dz} to model nonlinear evolution.

\section{Preliminaries}
\label{sec:preliminaries}

Gravitational lensing by large-scale structure preserves surface brightness, and therefore remaps the temperature anisotropy
and polarization of the CMB (for recent reviews see \cite{Lewis:2006fu,Hanson:2009kr}).  The remapping is described mathematically
by introducing a vector field ${\bf d}(\n)$ (the deflection field) such that lensed and unlensed CMB temperature and polarization
fields are related by:
\ba
T_{\rm len}(\n) &=& T_{\rm unl}(\n + {\bf d}(\n)) \\
(Q\pm iU)_{\rm len}(\n) &=& (Q\pm iU)_{\rm unl}(\n + {\bf d}(\n)) \,. %,
\ea
This equation is exact in the flat-sky limit, but schematic on the curved sky:
here the remapping $\n + {\bf d}$ represents an angular displacement 
$|{\bf d}|$ from $\n$ along a geodesic in the direction given by ${\bf d}$, 
and there is an implicit parallel transport of the $Q \pm iU$ pseudo-vector \cite{Challinor:2002cd}.
%where the ``+'' on the right-hand side of these equations should be interpreted as ordinary addition in the flat-sky limit,
%or parallel transport in the all-sky case \cite{Challinor:2002cd}.

To lowest order in perturbation theory, the deflection field ${\bf d}(\n)$ is the gradient of a scalar lensing potential (i.e.~${\bf d}(\n) = \nabla\phi(\n)$)
which can be written as a line-of-sight integral:
\be
\phi(\n) = -2 \int_0^{z_{\rm rec}} \frac{dz}{H(z)} \Psi(z, D(z)\n) \left( \frac{1}{D(z)} - \frac{1}{D(z_{\rm rec})} \right) \,,
\ee
where $H(z)$ is the Hubble factor, $\Psi(z,{\bf x})$ is the Newtonian potential, $D(z)$ denotes the comoving distance to redshift $z$, and $\Omega_K=0$ has been assumed.
In the Limber approximation, which we use throughout this paper, the angular power spectrum of the lensing potential is given by
\be
C_\ell^{\phi\phi} = 4 \int_0^{z_{\rm rec}} \frac{dz}{H(z) D(z)^2} P_\Psi(z,k=\ell/D(z)) \left( \frac{1}{D(z)} - \frac{1}{D(z_{\rm rec})} \right)^2 \,, \label{eq:clphi}
\ee
where $P_\Psi(z,k)$ 
is the power spectrum of the potential at redshift $z$.
The lensed B-mode $B^{\rm len}_{\ell_1m_1}$ is given in terms of the unlensed E-mode $E_{\ell_2m_2}$ at first order in the lensing potential
$\phi_{\ell m}$ by:
\be
B^{\rm len}_{\ell_1m_1} = \sum_{\ell_2m_2\ell m} f^{EB}_{\ell_1\ell_2\ell} \threej{\ell_1}{\ell_2}{\ell}{m_1}{m_2}{m} E^*_{\ell_2m_2} \phi^*_{\ell m} \,. \label{eq:Blen}
\ee
Here, the coupling coefficient $f^{EB}_{\ell_1\ell_2\ell}$ is given by
\be
f^{EB}_{\ell_1\ell_2\ell} = \frac{F^{-2}_{\ell_1\ell_2\ell} - F^2_{\ell_1\ell_2\ell}}{2i} \,,
\ee
where
\be
F^s_{\ell_1\ell_2\ell_3} \eqdef [-\ell_1(\ell_1+1) + \ell_2(\ell_2+1) + \ell_3(\ell_3+1)]
  \sqrt{\frac{(2\ell_1+1)(2\ell_2+1)(2\ell_3+1)}{16\pi}} \threej{\ell_1}{\ell_2}{\ell_3}{-s}{s}{0} \,. \label{eq:F}
\ee
A short calculation now shows that the power spectrum of the lensed B-mode is given by:
\be
C_{\ell_1}^{B_{\rm len}} = \frac{1}{2\ell_1+1} \sum_{\ell_2\ell} |f^{EB}_{\ell_1\ell_2\ell}|^2 C_{\ell_2}^{EE} C_{\ell}^{\phi\phi} \,.  \label{eq:ClBlen}
\ee
In writing down the form~(\ref{eq:Blen}) of the lensed B-mode, we have made the so-called gradient approximation, i.e.~keeping
only the lowest order term in the lensing potential, 
but this has been shown to be an excellent approximation for B-modes with $\ell \lsim 2000$ \cite{Challinor:2005jy}.

\begin{figure}[!ht]
\centerline{\includegraphics[width=13cm]{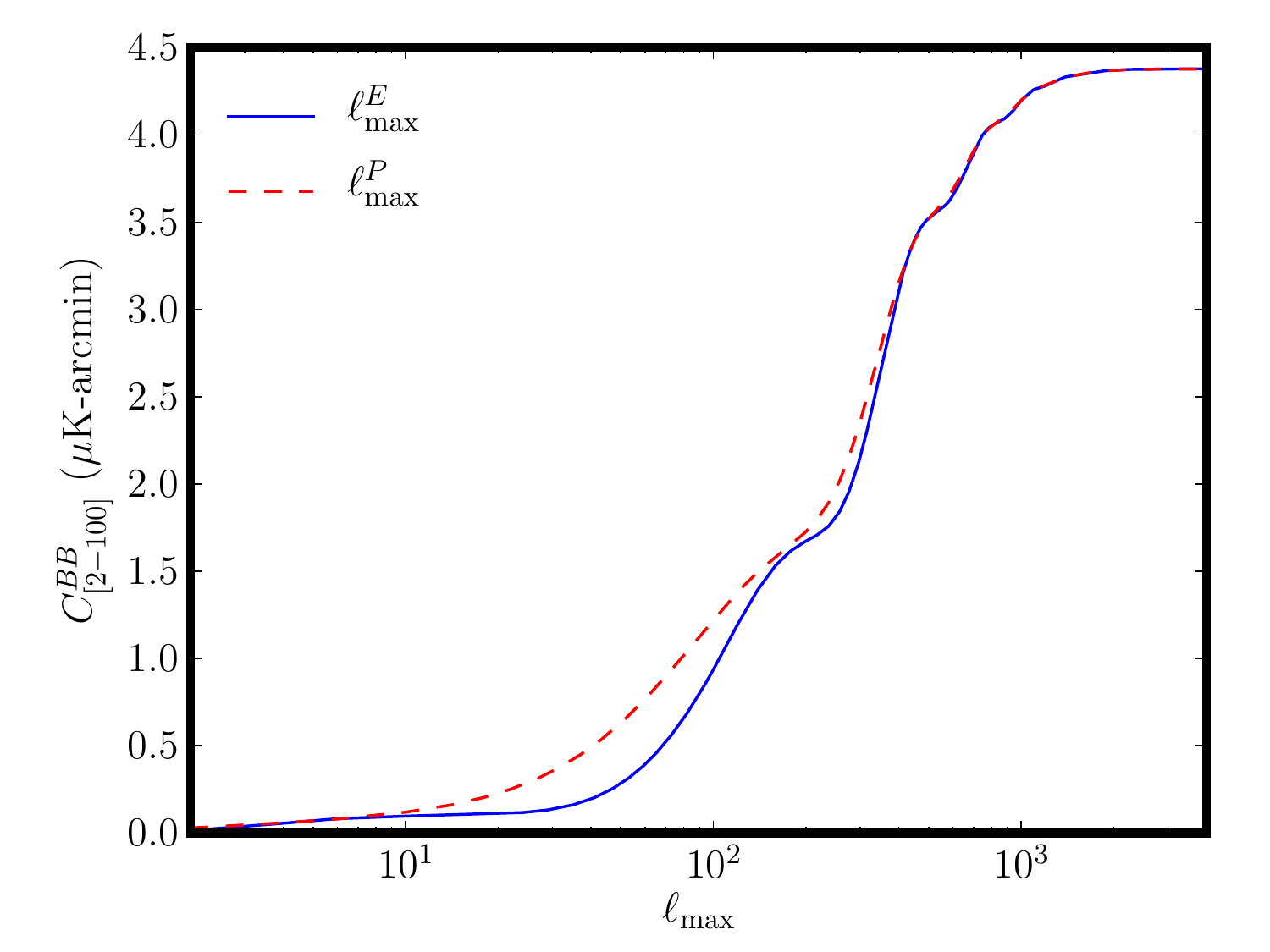}}
% \centerline{\epsfxsize=13cm\epsffile{clbb_sources.eps}}
\caption{Large scale lensing $B$-mode power spectrum, interpreted as a pixel noise level in $\mu$K-arcmin, for varying choices of maximum E-mode multipole
$\ellmax^E$ or deflection multipole $\ellmax^P$ in Eq.~(\ref{eq:ClBlen}).  It is seen that the large-scale B-mode is generated by E-mode and deflection power
on a wide range of angular scales (roughly $30 \lsim \ell \lsim 1000$).}
\label{fig:clbb_sources}
\end{figure}

To delens the observed (lensed) B-modes, two pieces of data are required: 
(1) a measurement $E^{\rm obs}_{\ell m}$ of the E-mode with noise power 
spectrum $N_\ell^{EE}$, 
and 
(2) a measurement $\phi^{\rm obs}_{\ell m}$ of the lensing potential $\phi$ with noise power spectrum $N_\ell^{\phi\phi}$.
Most commonly, the measurement of the lensing potential is obtained ``internally'' from the CMB, by applying
a lens reconstruction estimator, but the discussion in this section is general and would also apply to an ``external'' measurement
(e.g.~estimating $\phi$ from observations of large-scale structure).
One feature of delensing that we wish to emphasize is that
both measurements must go from intermediate to small angular scales (roughly $30 \lsim \ell \lsim 1000$), even if B-mode
delensing is only required on large scales, since small-scale lenses contribute large-scale B-mode power.
This is shown directly in Fig.~\ref{fig:clbb_sources}.

Intuitively, the idea of delensing is that a lensing-limited measurement $B^{\rm obs}_{\ell_1m_1}$ of the large-scale B-mode can be
improved by subtracting a quadratic combination of $E^{\rm obs}_{\ell_2m_2}$ and $\phi^{\rm obs}_{\ell m}$ which approximates the
lensed B-mode (Eq.~(\ref{eq:Blen})) as closely as possible.
More formally, the delensing estimator is given by
\be
B_{\ell_1 m_1}^{\rm del} = B_{\ell_1 m_1}^{\rm obs} - \sum_{\ell_2m_2\ell m} f^{EB}_{\ell_1\ell_2\ell} \threej{\ell_1}{\ell_2}{\ell}{m_1}{m_2}{m}
    \left( \frac{C_{\ell_2}^{EE} E^{\rm obs*}_{\ell_2m_2}}{C_{\ell_2}^{EE} + N_{\ell_2}^{EE}} \right)
    \left( \frac{C_\ell^{\phi\phi} \phi^{\rm obs*}_{\ell m}}{C_\ell^{\phi\phi} + N_\ell^{\phi\phi}} \right) \,.
\label{eq:Bdel}
\ee
If the power spectrum of the original B-mode measurement is the sum of contributions from tensor modes, lensing, and instrumental noise:
\be
C_\ell^{B_{\rm obs}} = C_\ell^{B_{\rm tens}} + C_\ell^{B_{\rm len}} + N_\ell^{BB} \,,
\ee
then the power spectrum of the delensed B-mode will be the sum of contributions from tensors, noise, and residual lensing:
\be
C_\ell^{B_{\rm del}} = C_\ell^{B_{\rm tens}} + C_\ell^{B_{\rm res}} + N_\ell^{BB} \,,
\ee
where the residual lensed B-mode power spectrum after delensing is given by:
\be
C_{\ell_1}^{B_{\rm res}} = \frac{1}{2\ell_1+1} \sum_{\ell_2\ell} |f^{EB}_{\ell_1\ell_2\ell}|^2 
  \left[ C_{\ell_2}^{EE} C_{\ell}^{\phi\phi} - 
              \left( \frac{ ( C_{\ell_2}^{EE} )^2 }{C_{\ell_2}^{EE}+N_{\ell_2}^{EE}} \right)
              \left( \frac{ ( C_{\ell}^{\phi\phi} )^2 }{C_{\ell}^{\phi\phi}+N_{\ell}^{\phi\phi}} \right)  \label{eq:ClBres}
  \right]
\ee
and satisfies $C_\ell^{B_{\rm res}} \le C_\ell^{B_{\rm len}}$.

The form~(\ref{eq:Bdel}) of the delensing estimator can be understood intuitively as Wiener filtering the observed E and $\phi$
fields, and using the filtered fields to estimate a lensed B-mode, which is then subtracted from the observed B-mode.
A formal derivation is obtained by solving for the weights on the right-hand side of~(\ref{eq:Bdel})
which minimize the residual power spectrum $C_\ell^{B_{\rm res}}$.  Details of this calculation are given
in Appendix~\ref{app:minimum_variance}.

On large angular scales ($\ell \lsim 100$), we find empirically that the residual lensing power spectrum $C_\ell^{B_{\rm res}}$
is independent of $\ell$ to an excellent approximation\footnote{To be more quantiative about this, $C_\ell^{B_{\rm res}}$ is not
observationally distinguishable (in the Fisher matrix sense) from a best-fit constant power spectrum, even with cosmic variance
limited all-sky measurements for $\ell\le 100$.} across the wide
range of instrumental specificiations considered in this paper.  This simplifies the interpretation and some implementational 
details of the forecasts which will follow: the residual lensed B-mode can be interpreted as a source of white noise, parameterized 
by a single number (the pixel noise in units $\mu$K-arcmin) which will depend on the instrumental specifications.
If no delensing is performed (i.e.~$C_\ell^{B_{\rm res}} = C_\ell^{B_{\rm len}}$), the pixel noise level is 4.4 $\mu$K-arcmin
in the fiducial cosmology from \S\ref{sec:intro}. % \kms{this paragraph needs another sentence or two.}

\section{Forecasts}

\begin{figure}[!ht]
\centerline{\includegraphics[width=13cm]{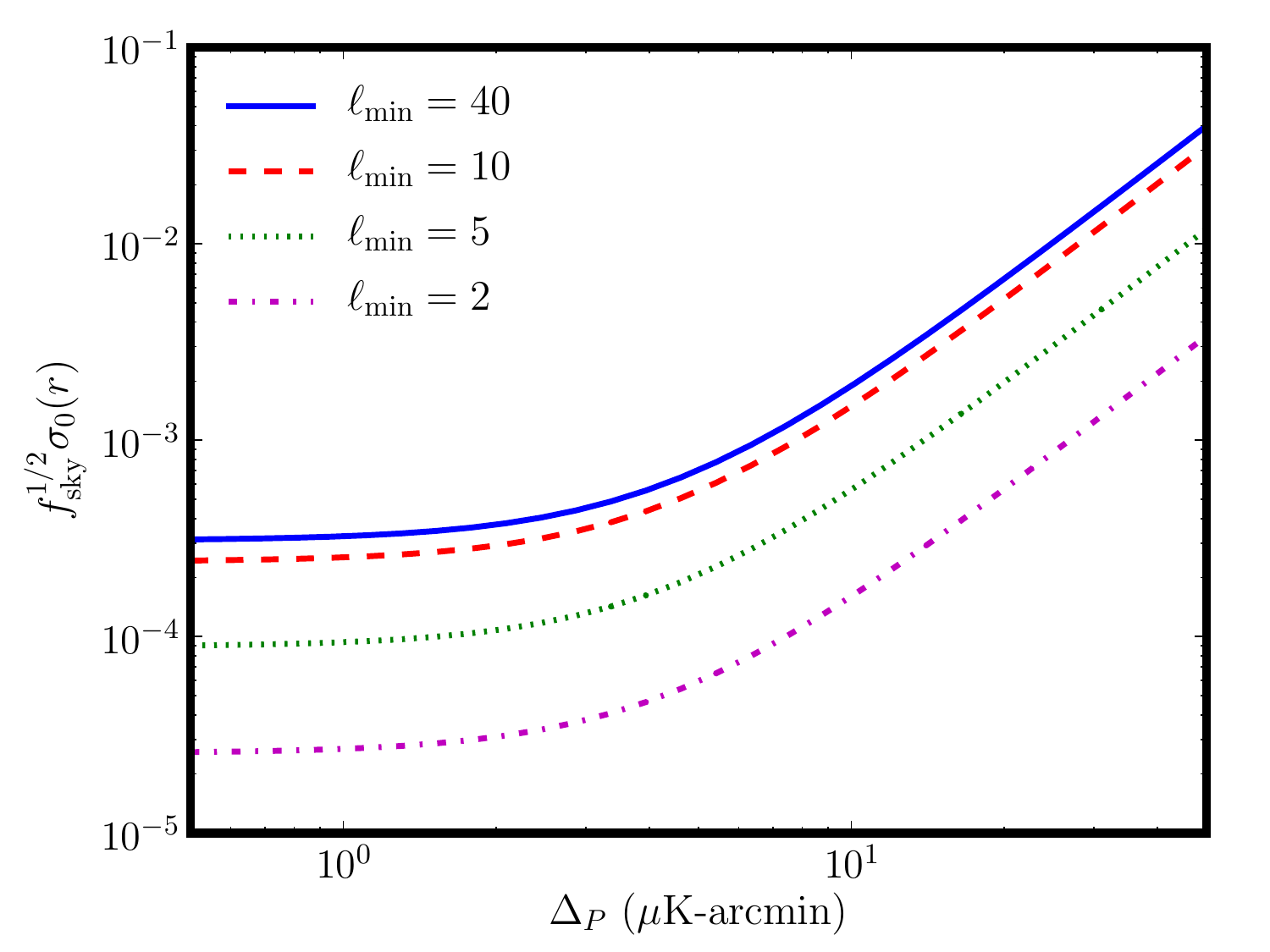}}
% \centerline{\epsfxsize=13cm\epsffile{sigma0.eps}}
\caption{Statistical error $\sigma_0(r)$ on the tensor-to-scalar ratio for varying noise level, assuming no delensing
has been done.  The forecast is strongly dependent on whether the reionization B-mode signal at $\ell\lsim 10$ is assumed
measureable in the presence of sky cuts and Galactic foregrounds.  The value of $\sigma_0(r)$ levels off for noise levels
$\lsim 4.4$ $\mu$K-arcmin since the observations have become lensing-limited.}
\label{fig:sigma0}
\end{figure}

In this paper, our basic figure of merit will be the statistical error $\sigma(r)$ on the tensor-to-scalar ratio $r$,
in the limit of no tensor modes
after delensing the large-scale B-mode, given by:
\be
\sigma(r) = \left[ \frac{\fsky}{2} \sum_{\ell\ge \ellmin} (2\ell+1) 
                \left( \frac{\partial C_\ell^{BB} / \partial r}{C_\ell^{B_{\rm res}} + N_\ell^{BB}} \right)^2
            \right]^{-1/2} \,. \label{eq:sigma}
\ee
Let $\sigma_0(r)$ denote the statistical error without performing delensing; it is given by replacing
$C_\ell^{B_{\rm res}} \rightarrow C_\ell^{B_{\rm len}}$ in the above expression:
\be
\sigma_0(r) = \left[ \frac{\fsky}{2} \sum_{\ell\ge\ellmin} (2\ell+1) 
                \left( \frac{\partial C_\ell^{BB} / \partial r}{C_\ell^{B_{\rm len}} + N_\ell^{BB}} \right)^2
            \right]^{-1/2} \,. \label{eq:sigma0}
\ee
We have included a parameter $\ellmin$ which represents the largest angular scale which can be measured in the presence
of a sky cut and large-scale foregrounds.  If this scale is large enough to include the signal at $\ell \lsim 10$ from 
tensor modes which enter the horizon during reionization, then the error on $r$ will be very sensitive to the value of $\ellmin$
which is assumed.  This is illustrated in Fig.~\ref{fig:sigma0}, which shows $\sigma_0(r)$ as a function of noise level for
a few choices of $\ellmin$.  For an ideal all-sky survey with $\ellmin=2$, the gravity wave measurement gets $\approx$90\% of 
its signal-to-noise from reionization scales ($\ell\lsim 10$), and $\approx$50\% of its signal-to-noise from the B-mode 
quadrupole ($\ell=2$) alone.  However, these large angular scales are very difficult to measure in suborbital experiments
(although not impossible with long-duration ballooning) and it is not clear whether the measuring the reionization signal
will be practical in the presence of foregrounds.  Although the reionization signal at $\ell\sim 8$ has a higher signal-to-noise
ratio than the recombination signal at $\ell\sim 60$, the signal-to-foreground ratio is expected to be higher for the
recombination signal \cite{Dunkley:2008am}, assuming that a patch of sky is observed which is chosen to minimize foreground
contamination.  For these reasons, it is currently unclear whether the reionization or recombination signal will offer
better observational prospects for constraining $r$.

As remarked in the previous section, the lensed B-mode power spectrum $C_\ell^{B_{\rm len}}$ and delensed power spectrum
$C_\ell^{B_{\rm res}}$ are independent of $\ell$ on large scales.  We will also assume that the instrumental beam is sufficiently
small that $N_\ell^{BB}$ is also approximately independent of $\ell$ on large scales.  In this case, if we define an improvement factor
$\alpha = \sigma_0(r)/\sigma(r)$, then $\alpha$ has the simple form:
\be
\alpha = \frac{\sigma_0(r)}{\sigma(r)} = \frac{C_\ell^{B_{\rm len}} + N_\ell^{BB}}{C_\ell^{B_{\rm res}} + N_\ell^{BB}} \,, \label{eq:alpha}
\ee
where the RHS is independent of $\ell$ (and the value of $\ellmin$ which is assumed).

Eq.~(\ref{eq:alpha}) states that $\sigma(r)$ factors into two pieces: a factor $\sigma_0(r)$ which depends mainly on
survey geometry (i.e.~$\fsky$ and $\ellmin$), and a factor $(1/\alpha)$ which represents the improvement due to delensing.\footnote{We
have shown this only for a ``sharp'' cutoff in $\ell$ parameterized by a cutoff multipole $\ellmin$, but it would also apply in the case
where the number of modes per multipole continuously drops to zero near the fundamental mode, e.g.~because ambiguous modes 
\cite{Bunn:2002df,Smith:2005gi} have been projected out.}
This is a very convenient simplification because it separates the issue of whether the reioinzation B-mode signal is measureable
from the issue of how well delensing improves the $r$ constraint.
In the following subsections, we will show forecasts for $\alpha$ given various observational scenarios, and remain agnostic
on the question of whether the reionization signal can be used to obtain a clean measurement of $r$.
The reader should bear in mind that the statistical error $\sigma(r)$ is given by $\sigma_0(r)/\alpha$, where the value of $\sigma_0(r)$
can be read off from Fig.~\ref{fig:sigma0} for a given survey geometry.

\subsection{Polarization delensing}
\label{ssec:polarization_delensing}

A B-mode measurement which extends to small angular scales ($\ell \gsim 200$) can be used to estimate the
lensing potential, by means of lens reconstruction estimators which cross-correlate E and 
B-modes \cite{Benabed:2000jt,Guzik:2000ju,Hu:2001kj,Hirata:2003ka}.
By taking this noisy estimate of $\phi$ as the input to the delensing estimator~(\ref{eq:Bdel}), delensing
can be performed in a way which is purely internal to a CMB polarization dataset.

We assume that the E and B-modes have been measured with instrumental noise power spectrum given by
\be
N_\ell^{EE} = N_\ell^{BB} = \Delta_P^2 \exp\left( \frac{\theta_{\rm FWHM}^2 \ell^2}{8 \log 2} \right) \,,
\ee
where $\Delta_P$ is the pixel noise level of the experiment (used with units $\mu$K-radian in the above
equation, but typically quoted in $\mu$K-arcmin).

The noise power spectrum of the lens reconstruction obtained using the EB estimator is given by \cite{Okamoto:2003zw}:
\be
N_\ell^{\phi\phi} = \left[ \frac{1}{2\ell+1} \sum_{\ell_1\ell_2} |f^{EB}_{\ell_1\ell_2\ell}|^2
    \left( \frac{1}{C_{\ell_1}^{BB}+N_{\ell_1}^{BB}} \right)
    \left( \frac{(C_{\ell_2}^{EE})^2}{C_{\ell_2}^{EE}+N_{\ell_2}^{EE}} \right)
    \right]^{-1} \,. \label{eq:nlphi_eb}
\ee
We can plug this into Eq.~(\ref{eq:ClBres}) to obtain $C_\ell^{B_{\rm res}}$, and then use Eq.~(\ref{eq:alpha})
to obtain a forecast for $\alpha$.
However, the resulting forecast is incomplete since it does not include the improvements that could be obtained
using iterative delensing.

In polarization, the quadratic estimators for the lensing potential and the delensed B-mode that have been
discussed so far can be significantly improved for low noise levels using an iterative, likelihood-based 
approach \cite{Hirata:2003ka}.
The intuition behind the iterative estimators is that lensed B-mode power acts as 
a source of noise for the lens reconstruction estimator (as can be seen from Eq.~(\ref{eq:nlphi_eb})), so that the
delensed B-mode (which has less power than the lensed B-mode) can be used as the input for a second round of delensing with improved
statistical errors, and so on iteratively.
One qualitative difference between the two estimators is that in the limit of zero instrumental noise ($N_\ell^{BB}=0$),
the iterative estimator can achieve perfect reconstruction of the lens potential $\phi$ and perfect delensing of the
B-mode.
As the instrumental noise improves, there is no fundamental limit to the value of $r$ which can be detected \cite{Seljak:2003pn},
unlike the case of the quadratic estimator \cite{Knox:2002pe,Kesden:2002ku}, although of course there will be some practical limit
due to foregrounds and instrumental systematics.

Based on this intuitive picture, we propose the following procedure for forecasting B-mode measurements which use iterative delensing.
After computing the residual lensing B-mode power spectrum $C_\ell^{B_{\rm res}}$ using Eqs.~(\ref{eq:nlphi_eb}),~(\ref{eq:ClBres})
as described above, we recompute $C_\ell^{B_{\rm res}}$ iteratively until convergence, using the value of $C_\ell^{B_{\rm res}}$ from
the previous iteration in place of $C_\ell^{B_{\rm len}}$.
To state this in a completely formal way, our forecasting procedure is to iterate the pair of equations
\ba
N_\ell^{\phi\phi} &=& \left[ \frac{1}{2\ell+1} \sum_{\ell_1\ell_2} |f^{EB}_{\ell_1\ell_2\ell}|^2
    \left( \frac{1}{C_{\ell_1}^{B_{\rm res}}+N_{\ell_1}^{BB}} \right)
    \left( \frac{(C_{\ell_2}^{EE})^2}{C_{\ell_2}^{EE}+N_{\ell_2}^{EE}} \right)
    \right]^{-1} \\
C_{\ell_1}^{B_{\rm res}} &=& \frac{1}{2\ell_1+1} \sum_{\ell_2\ell} |f^{EB}_{\ell_1\ell_2\ell}|^2 
  \left[ C_{\ell_2}^{EE} C_{\ell}^{\phi\phi} - 
              \left( \frac{ ( C_{\ell_2}^{EE} )^2 }{C_{\ell_2}^{EE}+N_{\ell_2}^{EE}} \right)
              \left( \frac{ ( C_{\ell}^{\phi\phi} )^2 }{C_{\ell}^{\phi\phi}+N_{\ell}^{\phi\phi}} \right)
  \right]
\ea
to convergence, starting by taking $C_\ell^{B_{\rm res}} = C_\ell^{B_{\rm len}}$ in the first iteration.

We have arrived at this forecasting procedure via a heuristic argument, but we can test its validity by comparing with the
results in Table~I of \cite{Seljak:2003pn}, which show values of $C_\ell^{B_{\rm res}}$ obtained from Monte Carlo
simulations of an iterative delensing estimator, for a wide range of instrumental parameters.
We find that all entries in the table agree at the $\approx$10\% level, showing that this simple heuristic procedure actually
provides rather accurate forecasts.  Given the implementational complexity and computational cost of the iterative
delensing estimator, this forecasting procedure is one of the main results of this paper.  Using the optimizations
from Appendix~\ref{app:real_space}, iterative delensing forecasts can be generated in a few CPU-seconds.

\begin{figure}[!ht]
\centerline{\includegraphics[width=13cm]{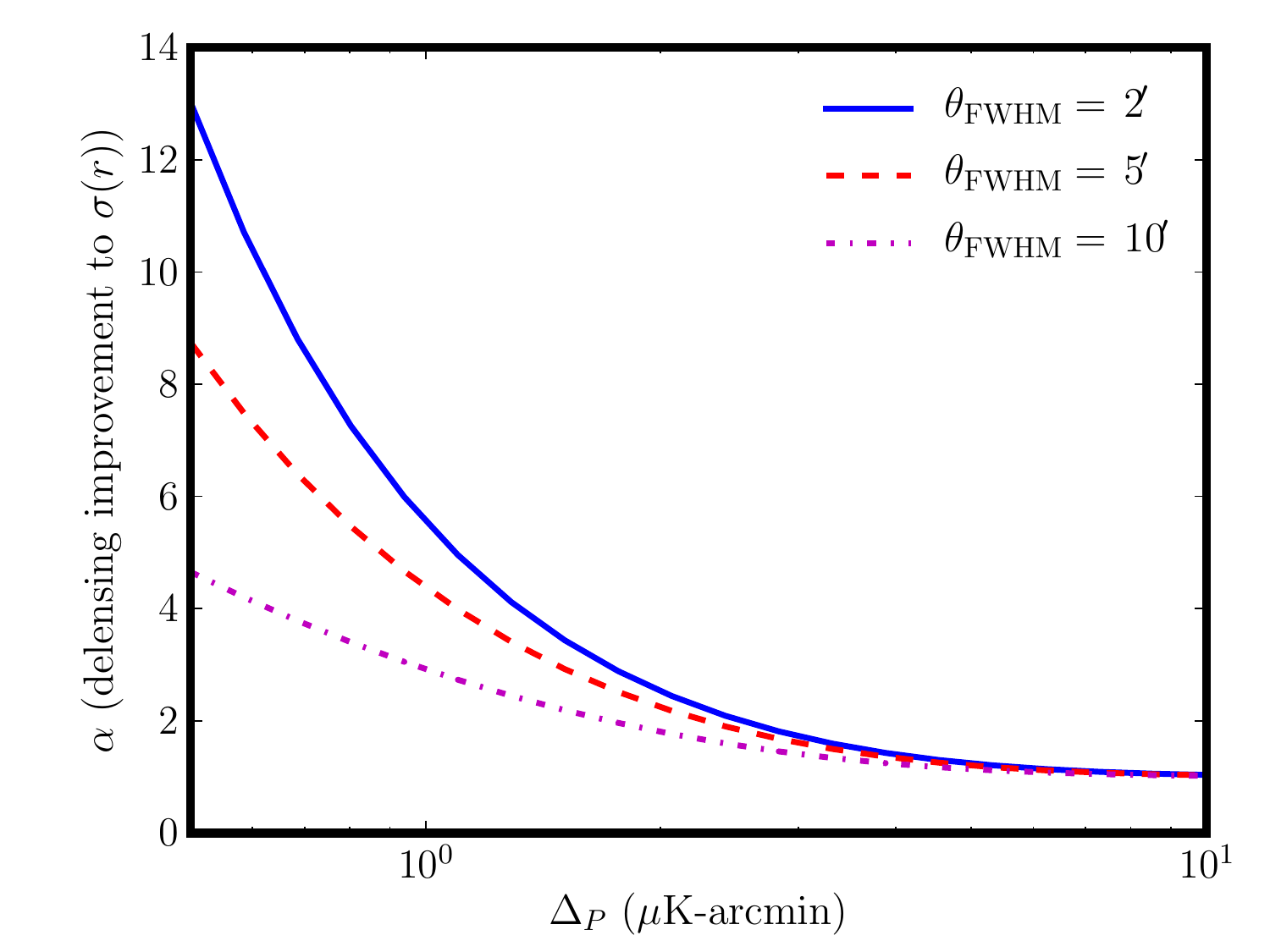}}
% \centerline{\epsfxsize=13cm\epsffile{alpha_pol.eps}}
\caption{Forecasted improvement $\alpha = \sigma_0(r)/\sigma(r)$ in the statistical error on $r$ due to polarization delensing,
for varying noise level and beam.  In the limit of low noise and high resolution, we find no limit (from delensing residuals
alone) to how well $r$ can be measured.}
\label{fig:alpha_pol}
\end{figure}

In Fig.~\ref{fig:alpha_pol}, we show forecasts for ``internal'' lens reconstruction using small-scale CMB polarization, 
for varying noise level and beam and taking $\ellmax=4000$ throughout.
Delensing can significantly improve $\sigma(r)$ if the instrumental noise is $\lsim 4.4$ $\mu$K-arcmin (so that the
large-scale B-mode is lensing-limited rather than noise limited) and the beam is small (so that the small-scale lensing
B-modes needed for lens reconstruction are measured).
As the instrumental noise and resolution improve, we do not find any fundamental limit (due to delensing residuals alone) to how well 
$r$ can be measured, in agreement with \cite{Seljak:2003pn}. % \kms{check this for noise levels lower than those shown in the plot?}

\subsection{Temperature delensing}

Lens reconstruction estimators can also be used to estimate the lensing potential from CMB {\em temperature} measurements 
which extend to small angular scales \cite{Bernardeau:1996aa,Bernardeau:1998mw,Zaldarriaga:1998te,Zaldarriaga:2000ud,
Hu:2001fa,Hu:2001tn,Cooray:2002py,Okamoto:2002ik,Hirata:2002jy,Kesden:2003cc,Hirata:2004rp}.
Lens reconstruction from temperature has been applied to detect CMB lensing in
WMAP, in cross-correlation with large-scale structure \cite{Smith:2007rg,Hirata:2008cb} and is expected to be a powerful source 
of cosmological information in the near future \cite{Hu:1996qs,Zaldarriaga:1997ch,Metcalf:1997ih,Stompor:1998zj,Hu:2000ti,Hu:2001bc,
Hu:2001fb,Abazajian:2002ck,Kaplinghat:2003bh,Lesgourgues:2005yv,Perotto:2006rj,dePutter:2009kn}.
It is therefore natural to ask whether the lens reconstruction from {\em temperature} datasets is sufficient to delens large-scale
CMB {\em polarization}.
Such a question would be relevant in a scenario where a lensing-limited B-mode measurement has been made on large angular scales,
using an instrumental beam which is too large to observe the small-scale lensing B-mode needed for ``internal'' delensing.
Note that, even if an ``external'' estimate of $\phi$ is available from CMB temperature, polarization delensing still requires a
measurement of E-mode polarization on intermediate to small scales (since the delensing estimator~(\ref{eq:Bdel}) combines E and $\phi$
on a wide range of scales, in order to to delens B on large scales).
Therefore, it is not possible to avoid the use of small-scale polarization completely, but the small-scale E-mode measurement is much
easier than the small-scale B-mode measurement needed for polarization delensing.  (One possible source of the small-scale E-mode measurement
would be the Planck dataset, which will release small-scale E-mode measurements on the full sky in the near future.)

Since CMB lens reconstruction estimators are based on the presence of small deviations from Gaussian statistics induced by lensing,
any non-Gaussian contribution to the CMB temperature can potentially bias lens reconstruction.
Although a detailed characterization of such biases is largely unknown territory, some studies have already found significant
biases from non-Gaussian signals such as the kinetic SZ effect \cite{Amblard:2004ih}.
Therefore, the limiting factor in temperature lens reconstruction is likely to be the presence of non-Gaussian secondary anisotropies
(which become increasingly important as $\ell$ increases), rather than instrumental sensitivity or resolution.
However, at the time of this writing it is unclear what range of scales will be ``sufficiently Gaussian'' to use for lens
reconstruction in practice.
We will model this unclear situation in a rough way by introducing a cutoff multipole $\ellmax^T$, and assuming that temperature multipoles
$\ell \le \ellmax^T$ can be used for lens reconstruction with the full statistical power of a Gaussian field (i.e. without introducing extra 
systematic error from secondary anisotropies), but multipoles $\ell > \ellmax^T$ are not used.
We will furthermore assume that the measurements for $\ell \le \ellmax^T$ have been measured with negligible instrumental noise.

\begin{figure}[!ht]
\centerline{\includegraphics[width=13cm]{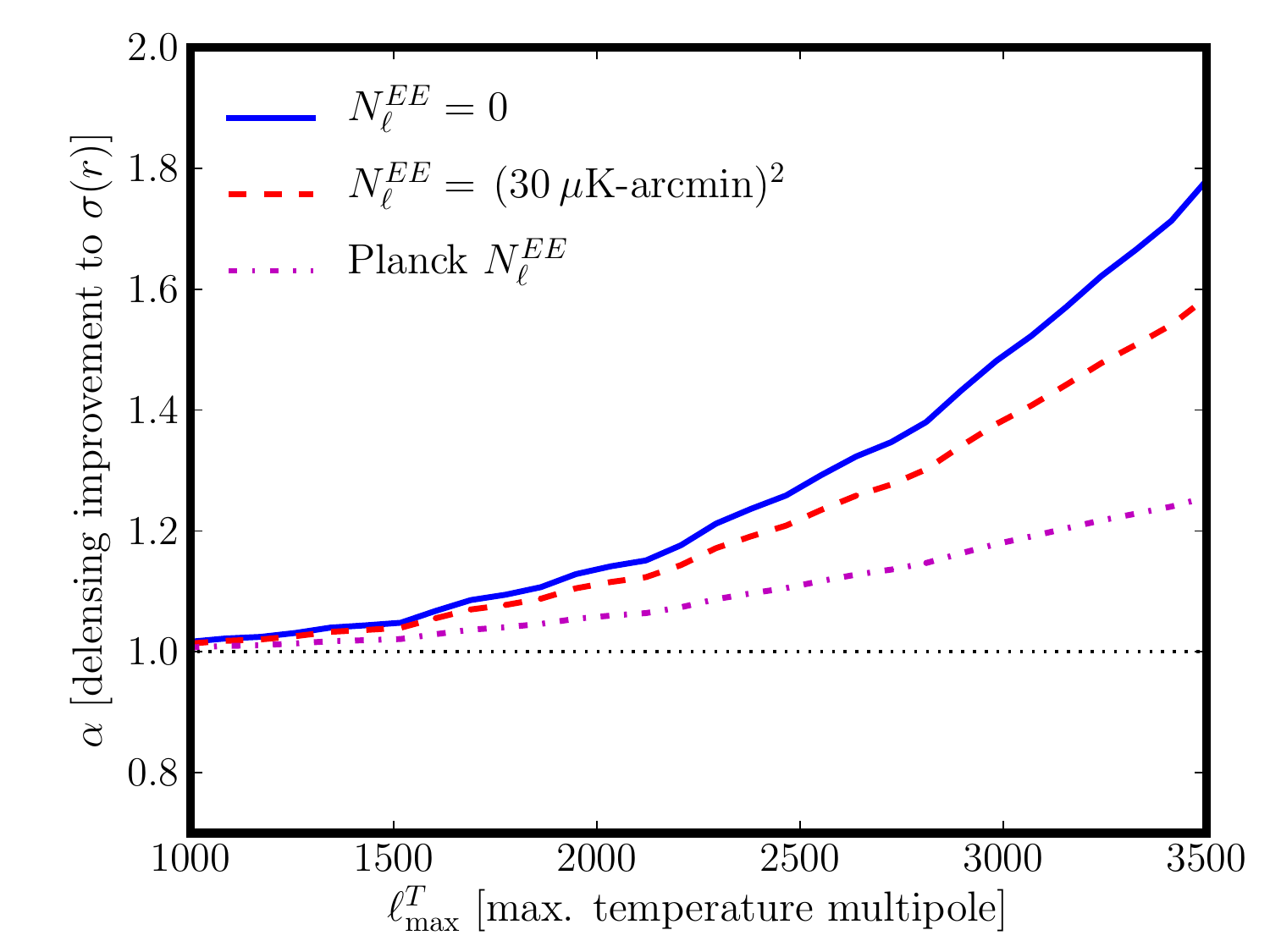}}
% \centerline{\epsfxsize=13cm\epsffile{alpha_t.eps}}
\caption{Forecasted improvement $\alpha = \sigma_0(r)/\sigma(r)$ in the statistical error on $r$, assuming a lensing-limited large-scale B-mode 
which is delensed using a lens reconstruction from small-scale {\em temperature}.  We assume that the temperature measurements are cosmic-variance limited
up to maximum multipole $\ellmax^T$ (beyond which non-Gaussian secondary anisotropy inhibits lens reconstruction).  Large-scale B-mode
delensing also requires a measurement of the small-scale E-mode; we consider either the E-mode measurement expected from Planck, a future
E-mode measurement with $\Delta_P=30$ $\mu$K-arcmin and zero beam, or a perfect E-mode measurement.}
\label{fig:alpha_t}
\end{figure}

Under these assumptions, the noise power spectrum of the lens reconstruction obtained using the TT estimator is given by \cite{Okamoto:2003zw}:
\be
N_\ell^{TT} = \left[ \frac{1}{2(2\ell+1)} \sum_{\ell_1,\ell_2 \le\ellmax^T} 
  \frac{ ( F^0_{\ell_1\ell_2\ell} C_{\ell_2}^{TT} + F^0_{\ell_2\ell_1\ell} C_{\ell_1}^{TT} )^2 }{ (C_{\ell_1}^{TT}+N_{\ell_1}^{TT})(C_{\ell_2}^{TT}+N_{\ell_2}^{TT}) }
\right]^{-1} \,.   \label{eq:nlphi_tt}
\ee
To obtain a delensing forecast, we must also specify two pieces of data:
\begin{enumerate}
\item The noise power spectrum $N_\ell^{BB}$ on large scales (needed to forecast $\alpha$ in Eq.~(\ref{eq:alpha})),
\item The noise power spectrum $N_\ell^{EE}$ on intermediate to small scales (needed to compute $C_\ell^{B_{\rm res}}$
in Eq.~(\ref{eq:ClBres})).
\end{enumerate}
For the first of these, we will take $N_\ell^{BB}=0$ in order to assume that the large-scale B-mode is purely lensing-limited.
For the second, we will consider three possibilities: either (a) Planck noise levels\footnote{When making forecasts which include
Planck E-mode measurements, we use the parameters from \cite{Albrecht:2006um}: we assume that the 143 GHz channel ($\Delta_P = 78$
$\mu$K-arcmin, $\theta_{\rm FWHM}=7.1$ arcmin) and the 217 GHz channel ($\Delta_P = 135$ $\mu$K-arcmin, $\theta_{\rm FWHM} = 5$ arcmin)
can be used with full sensitivity, while channels at lower or higher frequency are ``consumed'' by foreground cleaning.},
(b) E-mode measurements with $\Delta_P=30$ $\mu$K-arcmin and zero beam, and (c) E-mode measurements with zero
noise.  Option (c) is artificial, since a zero-noise measurement of the small-scale E-modes would be accompanied by a zero-noise measurement
of the lensing B-modes and one could just do polarization delensing, but we include it for the sake of having an upper bound on
the possible improvement from temperature delensing, by neglecting uncertainty in the small-scale E-mode.

\begin{figure}[!ht]
\centerline{\includegraphics[width=13cm]{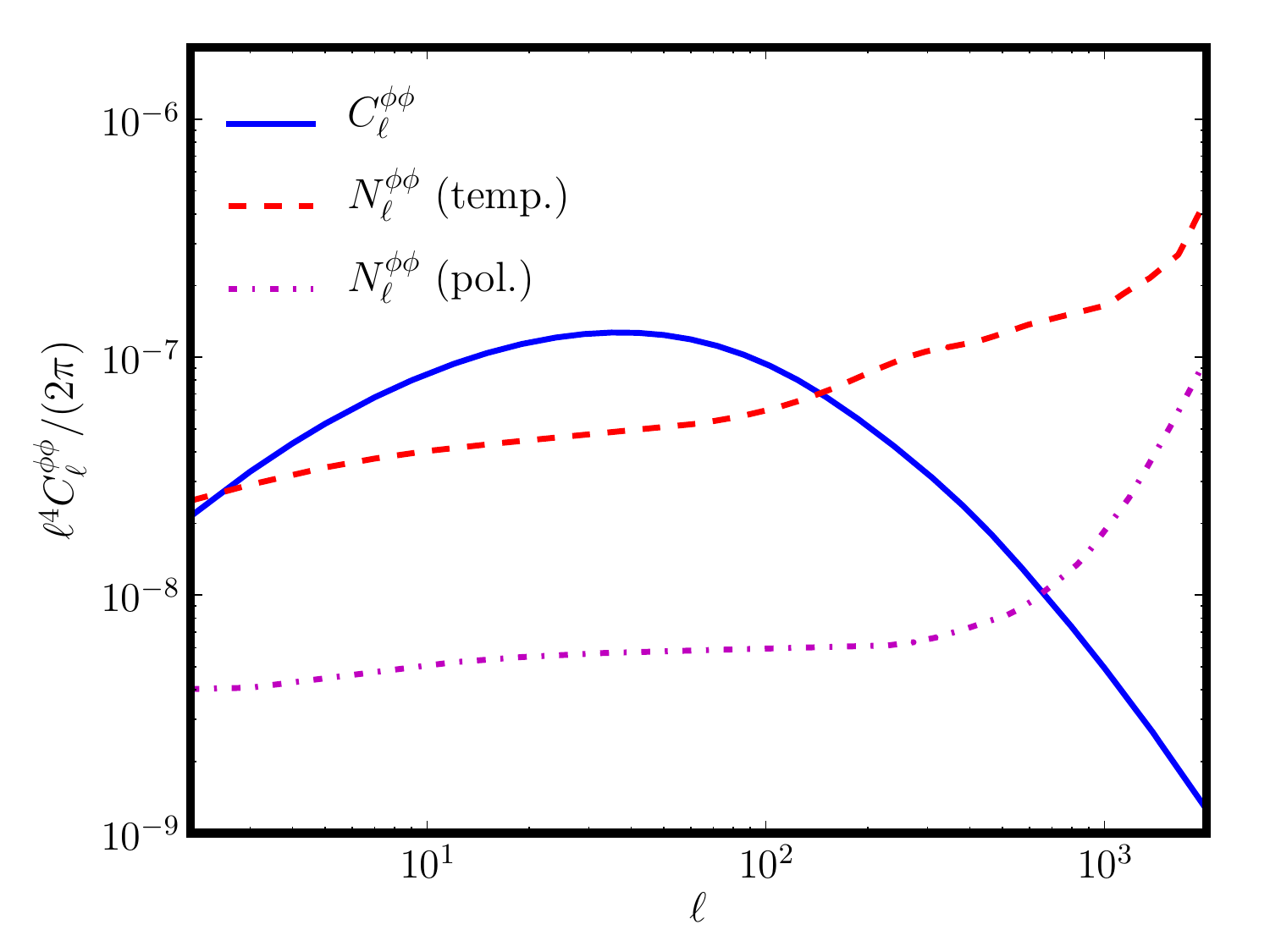}}
% \centerline{\epsfxsize=13cm\epsffile{nlphi.eps}}
\caption{Reconstruction noise power spectra $N_\ell^{\phi\phi}$ resulting from (a) temperature lens reconstruction assuming noise-free measurements to
maximum multipole $\ellmax^T=2500$, (b) polarization lens reconstruction with $\Delta_P=2$ $\mu$K-arcmin and $\theta_{\rm FWHM}=5$ arcmin.  The signal
power spectrum $C_\ell^{\phi\phi}$ is shown for comparison.}
\label{fig:nlphi}
\end{figure}

In Fig.~\ref{fig:alpha_t}, we show the improvement in $\sigma(r)$ that can be obtained from temperature delensing, for varying
$\ellmax^T$ and small-scale E-mode noise.
It is seen that the improvement in $\sigma(r)$ due to delensing is always modest, if the small-scale CMB temperature is the source
of the lens reconstruction.  Obtaining an improvement larger than $\approx$10\% would require both measurements of the small-scale E-mode 
which are less noisy than Planck, and making the optimistic assumption that temperature multipoles out to $\ellmax^T \approx 3000$ can be
cleaned of secondary anisotropy at a level where they can be used for lens reconstruction.
This negative result can be understood in a more qualitative way by comparing the noise power spectra $N_\ell^{\phi\phi}$ obtained from
temperature and polarization lens reconstruction (Fig.~\ref{fig:nlphi}).  The lensing potential on large scales ($\ell \lsim 100$) is
reconstructed with high signal-to-noise in both cases, but reconstructing lenses on small scales ($\ell\sim 1000$) requires polarization.
Since small-scale lenses do generate B-mode power on large scales (by coupling to small-scale E-modes), the temperature reconstruction is of
limited effectiveness in delensing the large-scale B-mode.

\subsection{Delensing using large-scale structure}

Finally, one could ask whether an ``external'' estimate of $\phi$ obtained from large-scale structure observations
can be used to delens large-scale B-mode polarization.
One could imagine using different flavors of large-scale structure data (for example, cosmic shear \cite{Marian:2007sr} or 21-cm temperature 
\cite{Zahn:2005ap,Sigurdson:2005cp,Mandel:2005xh}), with redshift and $\ell$ weighting optimized to approximate the CMB lensing potential $\phi$ as
closely as possible.
The problem encountered in the previous subsection, namely that the lens reconstruction from CMB temperature does not extend to small scales,
is not generally an issue for large-scale structure, which can go to even smaller scales than the CMB polarization reconstruction.
However, a new problem arises: the CMB lensing potential receives contributions from matter fluctuations at high redshift, and it is not possible
to include these contributions if $\phi$ is being estimated from large-scale structure data at low $z$.

Let us make the crude assumption that the matter fluctuations have been observed perfectly for redshifts $z\le\zmax$, and have not
been observed at all for $z > \zmax$.  (In a real survey, there will be a smooth transition between redshifts at which the matter density field
is signal-dominated versus noise-dominated, but we will use a sharp cutoff to get a rough idea of what to expect.)

A small modification to Eq.~(\ref{eq:ClBres}) for forecasting $C_\ell^{B_{\rm res}}$ is needed in the large-scale structure case.
In deriving Eq.~(\ref{eq:ClBres}), we have assumed that the measurement $\phi^{\rm obs}_{\ell m}$ of the lensing potential is the
sum of the true potential and an uncorrelated noise term (i.e.~that $(\phi^{\rm obs}_{\ell m} - \phi_{\ell m})$ and $\phi_{\ell m}$
are uncorrelated fields).
In the large-scale structure case, we write the true potential as the sum of uncorrelated contributions from low and high redshifts: 
$\phi_{\ell m} = \phi^{\rm lo}_{\ell m} + \phi^{\rm hi}_{\ell m}$.  We then assume that $\phi^{\rm lo}_{\ell m}$ has been observed
perfectly, that $\phi^{\rm hi}_{\ell m}$ is unobserved, and that we have an observation $E^{\rm obs}_{\ell m}$ of the small-scale
E-mode with noise power spectrum $N_\ell^{EE}$.
Given this setup, the minimum-variance delensing estimator and residual lensing B-mode power spectrum are given by:
\ba
B_{\ell_1 m_1}^{\rm del} &=& B_{\ell_1 m_1}^{\rm obs} - \sum_{\ell_2m_2\ell m} f^{EB}_{\ell_1\ell_2\ell}
    \left( \frac{C_{\ell_2}^{EE} E^{\rm obs*}_{\ell_2m_2}}{C_{\ell_2}^{EE} + N_{\ell_2}^{EE}} \right)
    \phi^{\rm lo*}_{\ell m} \label{eq:Bdel_lss} \\
C_{\ell_1}^{B_{\rm res}} &=& \frac{1}{2\ell_1+1} \sum_{\ell_2\ell} |f^{EB}_{\ell_1\ell_2\ell}|^2 
  \left[ C_{\ell_2}^{EE} C_{\ell}^{\phi\phi} - 
              \left( \frac{ ( C_{\ell_2}^{EE} )^2 }{C_{\ell_2}^{EE}+N_{\ell_2}^{EE}} \right) C_\ell^{\phi_{\rm lo}}
  \right] \,. \label{eq:ClBres_lss}
\ea
This is shown in Appendix~\ref{app:minimum_variance}, although the estimator~(\ref{eq:Bdel_lss}) is easy to guess: we simply subtract
an estimate for the lensed B-mode which is obtained by combining the Wiener-filtered E-mode and the observed part of the lensing potential.
Note that the power spectrum $C_\ell^{\phi_{\rm lo}}$ can be calculated in the Limber approximation by simply restricting the line-of-sight 
integral~(\ref{eq:clphi}) to the redshift range $0 \le z \le \zmax$.

\begin{figure}[!ht]
\centerline{\includegraphics[width=13cm]{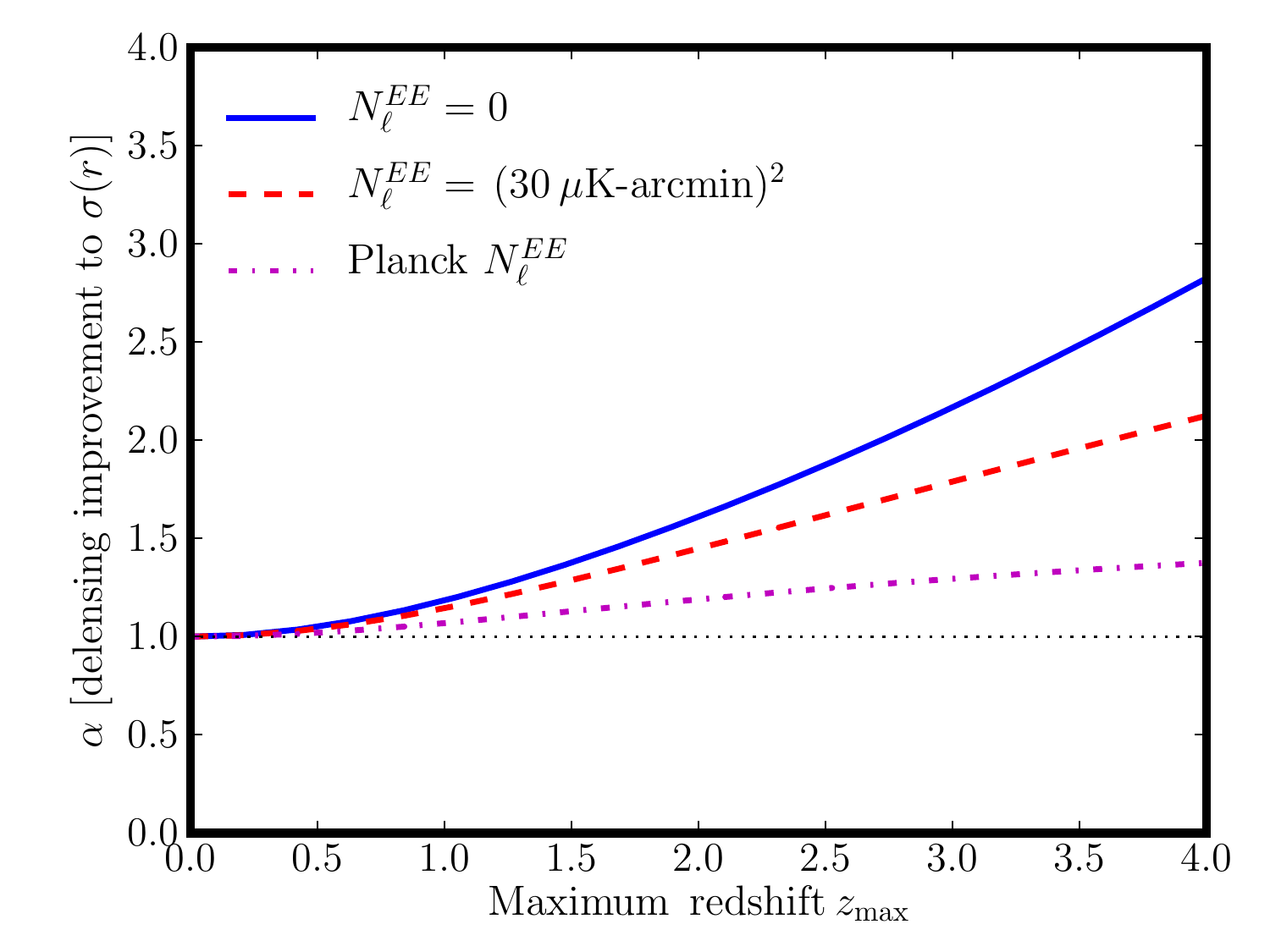}}
% \centerline{\epsfxsize=13cm\epsffile{alpha_lss.eps}}
\caption{Forecasted improvement $\alpha = \sigma_0(r)/\sigma(r)$ in the statistical error on $r$, assuming a lensing-limited large-scale
B-mode which is delensed using perfect large-scale structure measurements out to redshift $\zmax$.
Large-scale B-mode delensing also requires a measurement of the small-scale E-mode; we consider either the E-mode measurement expected from Planck, 
a future E-mode measurement with $\Delta_P=30$ $\mu$K-arcmin and zero beam, or a perfect E-mode measurement.}
\label{fig:alpha_lss}
\end{figure}

In Fig.~\ref{fig:alpha_lss}, we show the improvement in $\sigma(r)$ that can be obtained from large-scale structure delensing,
for varying $\zmax$.  As in the temperature case from the previous subsection, this type of delensing also requires a measurement
of the small-scale E-mode polarization, and we consider the same three cases: a Planck E-mode measurement, a measurement with 
$\Delta_P=30$ $\mu$K-arcmin, and a zero-noise measurement.
It is seen that the improvement in $\sigma(r)$ due to large-scale structure delensing is modest, at least for survey parameters
which are likely to be attainable for the next few generations of CMB polarization experiments.
An ambitious galaxy/quasar survey which is deep enough to probe structure to $z\sim 3$ with high signal-to-noise can delens by
a factor $\sim$2, in combination with small-scale E-mode measurements which are a few times more sensitive than Planck.

Qualitatively similar results were obtained in \cite{Marian:2007sr}, in a detailed analysis which concentrated on cosmic shear
as the large-scale structure observable.  Here, we have shown that these conclusions are generic to any observable which does
not probe the matter distributions at high redshift.  Even with perfect knowlege of the matter fluctuations out to redshift
$\sim$3, only relatively modest improvements in $\sigma(r)$ can be obtained.

\section{Discussion}

In this paper, we have presented forecasts for delensing the gravity wave B-mode, aimed at assessing whether high-sensitivity
measurements of CMB polarization on small scales are a requirement, or whether the lensing map can come from a different source.
We have also presented a simple framework for including delensing in forecasts, which is approximate but agrees very well with
Monte Carlo simulations.

One technical but important point is that calculations throughout this paper have treated all fields as Gaussian, even ``composite''
fields which are formed from products of others (such as the lensed B-mode, which is a composite of $E$ and $\phi$).
In more detail, we have made three approximations as follows.
First, the Fisher forecasts for $\sigma(r)$ and $\sigma_0(r)$ (Eqs.~(\ref{eq:sigma}),~(\ref{eq:sigma0})) have assumed that the bandpower
covariance of the lensed B-mode $B^{\rm len}_{\ell m}$ and delensed B-mode $B^{\rm res}_{\ell m}$ are related to the power spectrum
in the same way as for a Gaussian field.
For the lensed B-mode, it has been shown \cite{Smith:2004up} that this is a good approximation for narrow bandpowers, but breaks down 
for bandpowers wider than the coherence scale of the lenses ($\Delta\ell \gsim 200$).
Since the effective bandpower for constraining $r$ is narrow ($\Delta\ell \sim 40$), the approximation should be accurate in this case.
It seems plausible that the same qualitative statements are true for the delensed B-mode, but we defer investigation of this issue to
further work.
The second approximation appears in Eq.~(\ref{eq:ClBres}): when calculating the residual B-mode power spectrum $C_\ell^{B_{\rm res}}$,
we have used Wick's theorem to compute expectation values of products of the fields $B^{\rm unl}$, $E^{\rm obs}$, $\phi^{\rm obs}$.
More precisely, we have neglected the $(B^{\rm unl} E^{\rm obs} \phi^{\rm obs})$ three-point function and the $(E^{\rm obs} E^{\rm obs} \phi^{\rm obs} \phi^{\rm obs})$ 
connected four-point function.
The validity of this approximation is unclear in the case where $\phi^{\rm obs}$ is obtained by applying a quadratic lens reconstruction
estimator to the CMB; in this case the 3-point and 4-point terms which have been neglected will ``expand'' to a nontrivial CMB 4-point or 6-point
function.
The situation becomes even murkier if the delensing is iterative, since iteration will lead to $N$-point functions of even higher order.
The third approximation we have made is treating the lensing potential $\phi$ as a Gaussian field.
We have included nonlinear corrections to the power spectrum $C_\ell^{\phi\phi}$, but nonlinear evolution also generates higher
$N$-point functions which have been neglected.
A complete study of these approximations and potential biases is outside the scope of this paper, and will be pursued separately \cite{HSD}.
As remarked in \S\ref{ssec:polarization_delensing}, the close agreement between the forecasts in this paper and the Monte Carlo results from 
\cite{Seljak:2003pn} suggests that the impact of these approximations is small in forecasts, but may be important in actual data analysis.

We considered the possibility of using the CMB lens reconstruction from small scale {\em temperature} to help delens large-scale polarization.
We found that this is not a promising strategy and that it will be difficult to achieve a reduction in $\sigma(r)$ which is better than $\approx 10$\%
with realistic assumptions about small-scale secondary temperature anisotropy.
The limiting factor is poor signal-to-noise in the TT lens reconstruction on small angular scales ($\ell\sim 1000$).

We also considered the possibility of using large-scale structure to help delens the CMB.
While the forecasts here are not especially encouraging, they are not as grim as the temperature delensing case.
At some point in the future, it may be interesting to obtain a modest (say 50\%) improvement in $\sigma(r)$ using large-scale structure, as a
proof of concept of the delensing idea.  (Large-scale structure delensing may also be easier in practice than polarization delensing, since there
are fewer systematics which will cross-correlate the two datasets.)
Going beyond this would require small-scale polarization measurements and polarization delensing.
Another futuristic possibility is a 21-cm experiment with sufficient sensitivity and resolution to reconstruct the lenses out to redshift $\sim$10
\cite{Zahn:2005ap,Sigurdson:2005cp,Hilbert:2007jda}, but this is unlikely to be available for the next few generations of CMB polarization experiments.
We conclude that polarization delensing is the only prospect in the forseeable future 
for probing the gravity wave signal significantly beyond the noise floor imposed by gravitational lensing.

\subsection*{Acknowledgements}

This work was organized and initiated at the workshop ``CMB Polarization workshop: theory and foregrounds'' held at Fermilab from Jun 23--26 2008.
We would like to thank the organizers and staff for a stimulating and productive atmosphere.
A preliminary version of these results was included, with calculational details omitted, in the NASA Decadal Survey report \cite{Smith:2008an}.
KMS is supported by a Lyman Spitzer fellowship in the Department of Astrophysical Sciences at Princeton University.
ML is supported as a Friends of the Institute for Advanced Study Member and by the NSF though AST-0807444.
CH is supported by the U.S.~Department of Energy (DE-FG03-92-ER40701), the National Science Foundation (AST-0807337),
and the Alfred P.~Sloan Foundation.
OZ is supported by an Inaugural Fellowship by the Berkeley Center for Cosmological Physics.
This research was partly funded by NASA Mission Concept Study award NNX08AT71G S01.
We also acknowledge the organizational work of the Primordial Polarization Program Definition Team.

\bibliographystyle{ieeetr}
\bibliography{delensing_no_go}

\appendix

\section{Appendix: minimum-variance estimators}
\label{app:minimum_variance}

In this appendix, we derive the minimum-variance delensing estimator and the residual B-mode power spectrum,
in the lens reconstruction case (Eqs.~(\ref{eq:Bdel}),~(\ref{eq:ClBres})) and the large-scale structure case
(Eqs.~(\ref{eq:Bdel_lss}),~(\ref{eq:ClBres_lss})).

Considering first the lens reconstruction case, we assume that the observed B-mode is the sum of contributions from
tensor modes, noise, and lensing:
\ba
B^{\rm obs}_{\ell_1m_1} &=& B^{\rm tens}_{\ell_1m_1} + B^{\rm noise}_{\ell_1m_1} + B^{\rm len}_{\ell_1m_1} \nn \\
  &=& B^{\rm tens}_{\ell_1m_1} + B^{\rm noise}_{\ell_1m_1} + \sum_{\ell_2m_2\ell m} f^{EB}_{\ell_1\ell_2\ell} \threej{\ell_1}{\ell_2}{\ell}{m_1}{m_2}{m} E^*_{\ell_2m_2} \phi^*_{\ell m} \,,
\ea
and that we have noisy observations $E^{\rm obs}_{\ell_2m_2}$, $\phi^{\rm obs}_{\ell m}$ of the small-scale E-mode and lensing potential,
with noise power spectra $N_{\ell_2}^{EE}$ and $N_\ell^{\phi\phi}$.
We consider a general delensing estimator of the form
\ba
B^{\rm del}_{\ell_1m_1} = B^{\rm obs}_{\ell_1m_1} 
   - \sum_{\ell_2m_2\ell m} g_{\ell_1\ell_2\ell} \threej{\ell_1}{\ell_2}{\ell}{m_1}{m_2}{m} E^{\rm obs*}_{\ell_2m_2} \phi^{\rm obs*}_{\ell m}
\ea
with weights $g_{\ell_1\ell_2\ell}$ to be determined.  The form of the second term (in which $g$ does not depend on $m,m_1,m_2$) is the most general form which
is bilinear in $E$,$\phi$ and satisfies rotation invariance.
A short calculation shows that the power spectrum of the delensed B-mode is given by 
$C_\ell^{B_{\rm del}} = C_\ell^{B_{\rm tens}} + N_\ell^{BB} + C_\ell^{B_{\rm res}}$, where
\ba
C_{\ell_1}^{B_{\rm res}} &=& \frac{1}{2\ell_1+1} \sum_{\ell_2\ell} \Big[
  (f^{EB}_{\ell_1\ell_2\ell}f^{EB*}_{\ell_1\ell_2\ell} - f^{EB}_{\ell_1\ell_2\ell}g^*_{\ell_1\ell_2\ell} -
     g_{\ell_1\ell_2\ell}f^{EB*}_{\ell_1\ell_2\ell}) C_{\ell_2}^{EE} C_{\ell}^{\phi\phi} \nn \\
&& \hspace{2.5cm} + g_{\ell_1\ell_2\ell} g^*_{\ell_1\ell_2\ell} (C_{\ell_2}^{EE}+N_{\ell_2}^{EE}) (C_{\ell}^{\phi\phi} + N_{\ell}^{\phi\phi}) \Big] \,. \label{eq:mv1}
\ea
Now we determine $g_{\ell_1\ell_2\ell}$ by minimizing the residual power spectrum $C_\ell^{B_{\rm res}}$.  Differentiating~(\ref{eq:mv1}) with respect to $g$, we find
that the minimum occurs at
\be
g_{\ell_1\ell_2\ell} = f^{EB}_{\ell_1\ell_2\ell} \left( \frac{C_{\ell_2}^{EE}}{C_{\ell_2}^{EE}+N_{\ell_2}^{EE}} \right) 
  \left( \frac{C_\ell^{\phi\phi}}{C_\ell^{\phi\phi} + N_\ell^{\phi\phi}} \right)
\ee
and the value of $C_\ell^{B_{\rm res}}$ at minimum is given by
\be
C_{\ell_1}^{B_{\rm res}} = \frac{1}{2\ell_1+1} \sum_{\ell_2\ell} |f^{EB}_{\ell_1\ell_2\ell}|^2 
  \left[ C_{\ell_2}^{EE} C_{\ell}^{\phi\phi} - 
              \left( \frac{ (C_{\ell_2}^{EE})^2 }{C_{\ell_2}^{EE}+N_{\ell_2}^{EE}} \right)
              \left( \frac{ (C_{\ell}^{\phi\phi})^2 }{C_{\ell}^{\phi\phi}+N_{\ell}^{\phi\phi}} \right)
  \right] \,.
\ee
This completes the derivation of Eqs.~(\ref{eq:Bdel}),~(\ref{eq:ClBres}) and concludes the lens reconstruction case.

Moving on to the large-scale structure case, we assume that the lensing potential is a sum of uncorrelated terms
\be
\phi_{\ell m} = \phi^{\rm lo}_{\ell m} + \phi^{\rm hi}_{\ell m} \,,
\ee
where $\phi^{\rm lo}_{\ell m}$ has been observed without noise, and 
that we have a noisy observation $E^{\rm obs}_{\ell_2m_2}$ of the small-scale E-mode with noise power spectrum $N_\ell^{EE}$.

We consider the delensing estimator:
\ba
B^{\rm del}_{\ell_1m_1} = B^{\rm obs}_{\ell_1m_1} 
   - \sum_{\ell_2m_2\ell m} h_{\ell_1\ell_2\ell} \threej{\ell_1}{\ell_2}{\ell}{m_1}{m_2}{m} E^{\rm obs*}_{\ell_2m_2} \phi^{\rm lo*}_{\ell m} \,.
\ea
In this case, the residual B-mode power spectrum is given by
\ba
C_{\ell_1}^{B_{\rm res}} &=& \frac{1}{2\ell_1+1} \sum_{\ell_2\ell} \Big[
   f^{EB}_{\ell_1\ell_2\ell}f^{EB*}_{\ell_1\ell_2\ell} C_{\ell_2}^{EE} C_{\ell}^\phi
 + h_{\ell_1\ell_2\ell}h^*_{\ell_1\ell_2\ell} (C_{\ell_2}^{EE}+N_{\ell_2}^{EE}) C_{\ell}^{\phi_{\rm lo}} \nn \\
 && \hspace{2.5cm} - (f^{EB}_{\ell_1\ell_2\ell}h^*_{\ell_1\ell_2\ell} + h_{\ell_1\ell_2\ell}f^{EB*}_{\ell_1\ell_2\ell}) C_{\ell_2}^{EE} C_{\ell}^{\phi_{\rm lo}} \Big] \,.
\ea
Solving for the weights $h_{\ell_1\ell_2\ell}$ which minimize $C_\ell^{B_{\rm res}}$, we get
\be
h_{\ell_1\ell_2\ell} = f^{EB}_{\ell_1\ell_2\ell} \left( \frac{C_{\ell_2}^{EE}}{C_{\ell_2}^{EE}+N_{\ell_2}^{EE}} \right)
\ee
and residual B-mode power spectrum
\be
C_{\ell_1}^{B_{\rm res}} = \frac{1}{2\ell_1+1} \sum_{\ell_2\ell} |f^{EB}_{\ell_1\ell_2\ell}|^2 
  \left[ C_{\ell_2}^{EE} C_{\ell}^{\phi\phi} - 
              \left( \frac{ (C_{\ell_2}^{EE})^2 }{C_{\ell_2}^{EE}+N_{\ell_2}^{EE}} \right) C_\ell^{\phi_{\rm lo}}
  \right] \,.
\ee
This completes the derivation of Eqs.~(\ref{eq:Bdel_lss}),~(\ref{eq:ClBres_lss}) and finishes the large-scale structure case.

\section{Appendix: fast real-space expressions}
\label{app:real_space}

The forecasts in this paper require repeated calculations of the lensed B-mode power spectrum (Eq.~(\ref{eq:ClBlen}))
and lens reconstruction power spectra in the TT (Eq.~\ref{eq:nlphi_tt}) and EB (Eq.~(\ref{eq:nlphi_eb})) cases.
We have written expressions for these quantities as harmonic-space sums whose computational cost is $\bigoh(\ellmax^3)$.
In this appendix, we will present equivalent real-space expressions which can be evaluated exactly with cost $\bigoh(\ellmax^2)$.
The method to convert these expressions to real space was originally proposed in \cite{Dvorkin:2009ah}.
This optimization is very convenient since forecasts can be done in a few CPU-seconds for a given set of instrumental
specifications, permitting rapid exploration of parameter space.

We first note that the $F$ symbol defined previously in Eq.~(\ref{eq:F}) can be equivalently written
\ba
F^s_{\ell_1\ell_2\ell_3} &=& 
  -\sqrt{\frac{(2\ell_1+1)(2\ell_2+1)(\ell_3)(\ell_3+1)(2\ell_3+1)}{16\pi}} \nn \\
&& \hspace{0.25cm} \times \Bigg[    \sqrt{(\ell_2-s)(\ell_2+s+1)} \threej{\ell_1}{\ell_2}{\ell_3}{-s}{s+1}{-1} \nn \\
&& \hspace{1cm}  + \sqrt{(\ell_2+s)(\ell_2-s+1)} \threej{\ell_1}{\ell_2}{\ell_3}{-s}{s-1}{1} \Bigg] \,. \label{eq:Fmagic}
\ea
% \kms{Elaborate on this!}
We also have the identity
\be
\int_{-1}^1 d(\cos\theta) d^{\ell_1}_{s_1s_1'}(\theta) d^{\ell_2}_{s_2s_2'}(\theta) d^{\ell_3}_{s_3s_3'}(\theta) 
  = 2 \threej{\ell_1}{\ell_2}{\ell_3}{s_1}{s_2}{s_3} \threej{\ell_1}{\ell_2}{\ell_3}{s_1'}{s_2'}{s_3'} \,, \label{eq:wd3}
\ee
where $d^{\ell_i}_{s_is_i'}$ are Wigner $d$-functions such that $s_1+s_2+s_3 = s_1'+s_2'+s_3' = 0$.

Armed with identities~(\ref{eq:Fmagic}),~(\ref{eq:wd3}), a long but straightforward calculation shows that Eq.~(\ref{eq:ClBlen})
for the lensed B-mode power spectrum is equivalent to:
\ba
C_\ell^{B_{\rm len}} &=& \frac{\pi}{4} \int d(\cos\theta) \left( \zeta^E_{3,3}(\theta)\zeta^\phi_+(\theta) 
   + 2\zeta^E_{3,1}(\theta)\zeta^\phi_-(\theta) + \zeta^E_{1,1}(\theta)\zeta^\phi_+(\theta) \right) d^{\ell}_{22}(\theta) \nn \\
        && - \frac{\pi}{4} \int d(\cos\theta) \left( \zeta^E_{3,-3}(\theta)\zeta^\phi_-(\theta) 
   + 2\zeta^E_{3,-1}(\theta)\zeta^\phi_+(\theta) + \zeta^E_{1,-1}(\theta)\zeta^\phi_-(\theta) \right) d^{\ell}_{2,-2}(\theta) \,, \label{eq:ClBlen2}
\ea
where the correlation functions which appear are defined by
\ba
\zeta^E_{3,\pm 3}(\theta) &=& \sum_\ell \frac{2\ell+1}{4\pi} C_{\ell}^{EE} (\ell-2)(\ell+3) d^{\ell}_{3,\pm 3}(\theta) \label{eq:cf1} \\
\zeta^E_{3,\pm 1}(\theta) &=& \sum_\ell \frac{2\ell+1}{4\pi} C_{\ell}^{EE} \sqrt{(\ell-1)(\ell+2)(\ell-2)(\ell+3)} d^{\ell}_{3,\pm 1}(\theta) \\
\zeta^E_{1,\pm 1}(\theta) &=& \sum_\ell \frac{2\ell+1}{4\pi} C_{\ell}^{EE} (\ell-1)(\ell+2) d^{\ell}_{1,\pm 1}(\theta) \\
\zeta^\phi_\pm(\theta) &=& \sum_\ell \frac{2\ell+1}{4\pi} C_\ell^{\phi\phi} \ell(\ell+1) d^\ell_{1,\pm 1}(\theta) \,. \label{eq:cf2}
\ea
To obtain a fast algorithm for computing $C_\ell^{B_{\rm len}}$, we first note that the integral on the RHS of~(\ref{eq:ClBlen2}) can be evaluated 
exactly using Gauss-Legendre quadrature with $\lceil (3\ellmax+1)/2 \rceil$ points, since
the integrand is a polynomial of degree $\le 3\ellmax$.  We precompute the correlation functions in Eqs.~(\ref{eq:cf1})--(\ref{eq:cf2}) at the
Gauss-Legendre quadrature points with computational cost $\bigoh(\ellmax^2)$, evaluating the Wigner $d$-functions by upward recursion in $\ell$.
The integral can then be done for all values of $\ell$ with the same $\bigoh(\ellmax^2)$ cost, using the same recursion.
This completes the fast algorithm for $C_\ell^{B_{\rm len}}$.

Analogously, a fast algorithm for the EB lens reconstruction noise can be obtained using the expression (equivalent to Eq.~(\ref{eq:nlphi_eb})):
\ba
N_\ell^{\phi\phi} &=& \Bigg[ \frac{\pi}{4} \ell(\ell+1) \int d(\cos\theta) \left( \zeta^{E'}_{3,3}(\theta)\zeta^B_+(\theta)
                      - 2\zeta^{E'}_{3,-1}(\theta)\zeta^B_-(\theta) + \zeta^{E'}_{1,1}(\theta)\zeta^B_+(\theta) \right) d^\ell_{11}(\theta) \nn \\
   && \hspace{0.25cm} - \frac{\pi}{4} \ell(\ell+1) \int d(\cos\theta) \left( \zeta^{E'}_{3,-3}(\theta)\zeta^B_-(\theta) 
                      - 2\zeta^{E'}_{3,1}(\theta)\zeta^B_+(\theta) + \zeta^{E'}_{1,-1}(\theta) \zeta^B_-(\theta) \right) d^\ell_{1,-1}(\theta) \Bigg]^{-1} \,,
\ea
with correlation functions defined by
\ba
\zeta^{E'}_{3,\pm 3}(\theta) &=& \sum_\ell \frac{2\ell+1}{4\pi} \frac{(C_{\ell}^{EE})^2}{C_{\ell}^{EE}+N_{\ell}^{EE}} (\ell-2)(\ell+3) d^{\ell}_{3,\pm 3}(\theta) \\
\zeta^{E'}_{3,\pm 1}(\theta) &=& \sum_\ell \frac{2\ell+1}{4\pi} \frac{(C_{\ell}^{EE})^2}{C_{\ell}^{EE}+N_{\ell}^{EE}} \sqrt{(\ell-1)(\ell+2)(\ell-2)(\ell+3)} d^{\ell}_{3,\pm 1}(\theta) \\
\zeta^{E'}_{1,\pm 1}(\theta) &=& \sum_\ell \frac{2\ell+1}{4\pi} \frac{(C_{\ell}^{EE})^2}{C_{\ell}^{EE}+N_{\ell}^{EE}} (\ell-1)(\ell+2) d^{\ell}_{1,\pm 1}(\theta) \\
\zeta^B_\pm(\theta) &=& \sum_{\ell} \frac{2\ell+1}{4\pi} \frac{1}{C_{\ell}^{BB}+N_{\ell}^{BB}} d^\ell_{2,\pm 2}(\theta) \,.
\ea
For the TT lens reconstruction noise , we use the expression (equivalent to Eq.~(\ref{eq:nlphi_tt})):
\ba
N_\ell^{\phi\phi} &=& \Bigg[ \pi \ell(\ell+1) \int d(\cos\theta) \left( \zeta^T_{00}(\theta)\zeta^T_{11}(\theta) 
                  - \zeta^T_{01}(\theta)\zeta^T_{01}(\theta) \right) d^\ell_{-1,-1}(\theta) \nn  \\
             && + \pi \ell(\ell+1) \int d(\cos\theta) \left( \zeta^T_{00}(\theta)\zeta^T_{1,-1}(\theta)
                   - \zeta^T_{01}(\theta)\zeta^T_{0,-1}(\theta) \right) d^\ell_{1,-1}(\theta) \Bigg]^{-1} \,,
\ea
with correlation functions defined by
\ba
\zeta^T_{00}(\theta) &=& \sum_\ell \frac{2\ell+1}{4\pi} \frac{1}{C_\ell^{TT}+N_\ell^{TT}} d^\ell_{00}(\theta)  \\
\zeta^T_{0,\pm 1}(\theta) &=& \sum_\ell \frac{2\ell+1}{4\pi} \sqrt{\ell(\ell+1)} \frac{C_\ell^{TT}}{C_\ell^{TT}+N_\ell^{TT}} d^\ell_{0,\pm 1}(\theta)  \\
\zeta^T_{1,\pm 1}(\theta) &=& \sum_\ell \frac{2\ell+1}{4\pi} \ell(\ell+1) \frac{(C_\ell^{TT})^2}{C_\ell^{TT}+N_\ell^{TT}} d^\ell_{1,\pm 1}(\theta) \,.
\ea
Fast algorithms can also be given for lens reconstruction noise power spectra which arise from TE, TB, and EE estimators, but we omit these
cases here since they are not needed in this paper. 
Indeed, fast algorithms exist for calculating the noise power spectra
for many quadratic estimators of statistical anisotropy, provided that
the weights which they use are separable.

\end{document}